\documentclass[11pt]{llncs}

\usepackage[english]{babel}
\usepackage[T1]{fontenc}
\usepackage[latin1]{inputenc}
\usepackage{times}
\usepackage{amssymb,amsmath,amscd,latexsym,graphicx}
\usepackage{tikz}
\usepackage{framed}

\newcommand{\val}[1]{\mathrm{val}(#1)}

\newcommand{\D}{\mathcal{D}} 
\newcommand{\prob}[1]{\mathbb{P}_{#1}}

\newcommand{\A}{\mathcal{A}}

\title{What is known about the Value $1$ Problem\\
for Probabilistic Automata?}

\author{Nathana\"el Fijalkow}

\institute{LIAFA, Paris 7,\\
University of Warsaw}

\begin{document}

\maketitle

\begin{abstract}
The value $1$ problem is a decision problem for probabilistic automata over finite words: are there words accepted by the automaton with arbitrarily high probability?
Although undecidable, this problem attracted a lot of attention over the last few years.
The aim of this paper is to review and relate the results pertaining to the value $1$ problem.

In particular, several algorithms have been proposed to partially solve this problem.
We show the relations between them, leading to the following conclusion:
the Markov Monoid Algorithm is the \textit{most correct} algorithm known to (partially) solve the value 1 problem.
\end{abstract}

\section{Introduction}
In 1963 Rabin~\cite{Rabin63} introduced the notion of probabilistic
automata, which are finite automata with randomized transitions.
This powerful model has been widely studied and has applications in many
fields like image processing~\cite{CK97}, computational biology~\cite{DEKM99} and speech processing~\cite{M97}.
Several algorithmic properties of probabilistic automata have been
considered in the litterature.
For instance, Sch\"utzenberger~\cite{Schutzenberger61} proved in 1961 that \textit{functional equivalence} 
is decidable in polynomial time (see also~\cite{Tzeng92}),
and even faster with randomized algorithms, which led to applications in
software verification~\cite{KMOWW11}.

However, many natural decision problems are undecidable,
and part of the literature on probabilistic automata is about
\textit{undecidability results}.
For example the \textit{emptiness}, the \textit{isolation} and the \textit{value} $1$
problems are undecidable, as shown in~\cite{Paz71,Bertoni77,GO10}.
To overcome untractability results, a lot of effort went into finding subclasses of probabilistic automata
for which natural decision problems become decidable.
For instance, the papers~\cite{KVAK10,CKVAK11} look at restrictions implying a decidable model-checking problem against $\omega$-regular specifications, 
and the paper~\cite{CSV13} investigates whether assuming isolated cut-points leads to decidability for the emptiness problem.

\vskip1em
We focus here on the efforts made to understand the value $1$ problem.
The aim of this paper is to review and relate the attempts made in this direction
over the last few years~\cite{GO10,CSV11,FGO12,CT12,BBG12,FGKO14}.

\section{Definitions}
Let $Q$ be a finite set of states. 
A probability distribution over $Q$ is a function $\delta : Q \rightarrow [0,1]$
such that $\sum_{q \in Q} \delta(q) = 1$.

Let $A$ be a finite alphabet.
The transitions of a probabilistic automaton are given by a function 
$\Delta : Q \times A \rightarrow \D(Q)$;
equivalently, for each letter $a \in A$ we consider 
a probabilistic transition matrix $M_a$, which is a square matrix in $[0,1]^{Q \times Q}$
such that every row of $M_a$ is a probability distribution over $Q$.
The value of $M_a(s,t)$ is the probability to go from state $s$ to state $t$ 
when reading the letter $a$.

Given an input word $w \in A^*$,
we denote $\prob{\A}(s \xrightarrow{w} t)$ 
the probability to go from state $s$ to state $t$ when reading the word $w$.
Formally, if $w = a_1 a_2 \cdots a_n$ then 
$\prob{\A}(s \xrightarrow{w} t) = (M_{a_1} M_{a_2} \cdots M_{a_n})(s,t)$.

\begin{definition}[Probabilistic automaton]
A tuple $\A = (Q, A, q_0, \Delta, F)$ represents a probabilistic automaton,
where $Q$ is a finite set of states, $A$ is the finite input alphabet, 
$q_0 \in Q$ is the initial state,
$\Delta$ define the transitions
and $F \subseteq Q$ is the set of accepting states.
\end{definition}

\begin{definition}[Acceptance probability]
The \emph{acceptance probability} of a word $w \in A^*$ by $\A$ is
$\sum_{f \in F} \prob{\A}(q_0 \xrightarrow{w} f)$, denoted $\prob{\A}(w)$.
\end{definition}

\begin{definition}[Value]
The \emph{value} of $\A$, denoted $\val{\A}$,
is the supremum acceptance probability over all possible input words:
\begin{equation}
\label{eq:value}
\val{\A} = \sup_{w \in A^*} \prob{\A}(w)\enspace.
\end{equation}
\end{definition}

We are interested in the following decision problem:
\begin{framed}
Given a probabilistic automaton $\A$, decide whether $\val{\A} = 1$.
\end{framed}

\section{An Equivalent Formulation and the Exact Computational Complexity}
The first result about the value $1$ problem is its surprising undecidability,
obtained with an elementary proof by Hugo Gimbert and Youssouf Oualhadj in~\cite{GO10}.

In a related yet seemingly different line of work, Christel Baier, Marcus Gr\"o\ss er and Nathalie Bertrand
undertook a thorough study of probabilistic B\"uchi automata~\cite{BG05,BBG08,BBG09,BBG12}.
One of the results obtained there is the undecidability of the emptiness problem for probabilistic B\"uchi automata
with probable semantics. It turns out that the two problems are actually Turing-equivalent:
\begin{itemize}
	\item the value $1$ problem for probabilistic automata over finite words,
	\item the emptiness problem for probabilistic B\"uchi automata with probable semantics.
\end{itemize} 
A first (very simple) reduction has been explained in~\cite{BBG12}: from a probabilistic automaton $\A$ over finite words,
one can construct a probabilistic B\"uchi automaton $\A'$ of linear size, such that $\val{\A} = 1$ if and only if $\A'$ is non-empty for the probable semantics.
The converse reduction is more involved, and follows from~\cite{CSV13}, but here the constructed automaton is of exponential size.

Even better, the exact computational complexity has been given in~\cite{CSV13}: both problems are $\Sigma^0_2$-complete.

\begin{theorem}[\cite{BBG12,CSV13}]
The value $1$ problem for probabilistic automata over finite words and the emptiness problem for probabilistic B\"uchi automata with probable semantics
are Turing-equivalent and $\Sigma^0_2$-complete.
\end{theorem}

\section{Decidable Subclasses of Probabilistic Automata}

Several subclasses of probabilistic automata were constructed in order to decide the value $1$ problem on such instances.

The first class was the $\sharp$-acyclic automata by Gimbert and Oualhadj~\cite{GO10}.

Later but concurrently, two different works have been published in the very same conference.
The first one introduces simple automata and structurally simple automata, by Krishnendu Chatterjee and Mathieu Tracol~\cite{CT12}.
The second, by Hugo Gimbert, Youssouf Oualhadj and the author introduces leaktight automata~\cite{FGO12}.

Although geared towards the same goal (deciding the value $1$ problem), the two classes came from different perspectives.
The paper of Krishnendu Chatterjee and Mathieu Tracol relies on a theorem from Probability Theory, called
the jet decompositions of (infinite) Markov Chains.
The paper of Hugo Gimbert, Youssouf Oualhadj and the author relies on a theorem from Algebra, called Simon's theorem,
asserting the existence of factorization trees of bounded height.

\vskip1em
Subsequent studies~\cite{FGKO14} showed that the class of leaktight automata actually strictly contains all the other classes,
implying that the Markov Monoid Algorithm used to decide the value $1$ problem for leaktight automata
actually decides the value $1$ problem for all cases where it is known to be decidable.

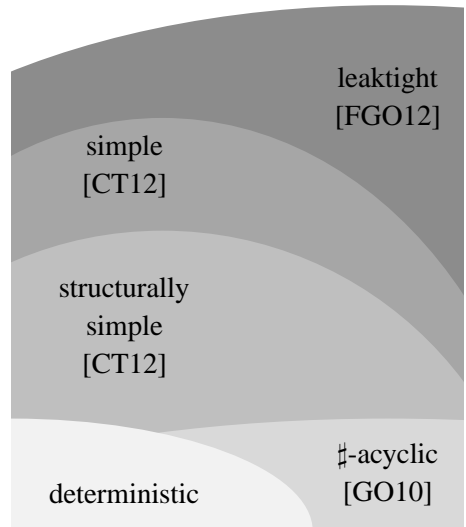
\begin{center}
\begin{tikzpicture}
\clip (-6,0) rectangle (0,8);

\fill[gray!90] (-1,3) ellipse (8cm and 4cm) ;
\draw (-1,6) node {leaktight} ;
\draw (-1,5.5) node {\cite{FGO12}} ;

\fill[gray!70] (-4,-1) ellipse (5.2cm and 6.5cm) ;
\draw (-4.5,5.1) node {simple} ;
\draw (-4.5,4.6) node {\cite{CT12}} ;

\fill[gray!50] (-4,-1) ellipse (5cm and 5cm) ;
\draw (-4.5,3.2) node {structurally} ;
\draw (-4.5,2.7) node {simple} ;
\draw (-4.5,2.2) node {\cite{CT12}} ;

\fill[gray!30] (-1,-1) ellipse (8cm and 2.5cm) ;
\draw (-1,1) node {$\sharp$-acyclic} ;
\draw (-1,.5) node {\cite{GO10}} ;

\fill[gray!10] (-6,0) ellipse (4cm and 1.5cm) ;
\draw (-4.5,.5) node {deterministic} ;
\end{tikzpicture}
\end{center}

\section*{Conclusion and Perspectives}
In this paper, we discussed some recent developments about the value $1$ problem.
We first gathered some results from the literature, explaining that it is actually Turing-equivalent
to the emptiness for probabilistic B\"uchi automata with the probable semantics, and
$\Sigma_2^0$-complete.
Then we presented the different attempts to decide the value $1$ problems on subclasses of probabilistic automata.
As a conclusion, the Markov Monoid Algorithm introduced in~\cite{FGO12}, used to decide the value $1$ problem
for leaktight automata, is actually the \textit{most correct} algorithm known so far,
as the class of leaktight automata strictly contains all other classes for which the value $1$ problem is known
to be decidable.

This motivates a deeper understanding of this algorithm. We know that the Markov Monoid Algorithm
cannot solve the value $1$ problem, as this problem is undecidable, but then what is the problem solved by this algorithm? 
In other words, can we characterize for which probabilistic automata the Markov Monoid Algorithm finds a value $1$ witness?

\bibliographystyle{alpha}
\bibliography{bib}

\end{document}